\documentclass[useAMS,usenatbib]{mn2e}

\usepackage{epsfig}
\usepackage{epstopdf}
\usepackage{lscape} 

\def\gtrsim{\mathrel{\hbox{\rlap{\hbox{\lower4pt\hbox{$\sim$}}}\hbox{$>$}}}}

\title[Interstellar Plasma Scattering]{Refractive Convergent Plasma Lenses explain ESE and pulsar scintillation}
\author[Pen et al]{Ue-Li
  Pen,$^{1}$\thanks{E-mail:\ pen@cita.utoronto.ca}
Lindsay King,$^2$\thanks{E-mail:\ ljk@ast.cam.ac.uk}
}
\begin{document}

\date{\today}

\pagerange{\pageref{firstpage}--\pageref{lastpage}} 
\pubyear{2011}

\maketitle
\label{firstpage}
\begin{abstract}
We propose convergent plasma lenses, possibly from current sheets, as
a generic solution to strong interstellar scattering.  These lenses
resolve the overpressure problem by geometric alignment as noted by
Goldreich and Shridhar (2006).  They further quantitatively explain
properties of extreme scattering events, and pulsar parabolic arcs.
This model makes quantitative predictions testable by VLBI on
scattering events.  It differs conceptually from previous models by
occurring through rare, 
localized underdense sheets.  Such sheets are thermally and
kinematically stable, and could be consequences of reconnection.  The
apparent diffractive effects are a result of coherent interference of
refractive images. We propose that these lenses can be used for
precision distance determination to pulsars, enabling accurate gravity
source localization.
\end{abstract}

\newcommand{\be}{\begin{eqnarray}}
\newcommand{\ee}{\end{eqnarray}}
\newcommand{\beq}{\begin{equation}}
\newcommand{\eeq}{\end{equation}}

\section{Introduction}

Several scintillation phenomena in the interstellar medium have posed
challenges to any physical model to explain
them\citep{2007ASPC..365..207R}.  These include: 
\\\noindent
(1) extreme scattering events (ESE) \citep{1987Natur.326..675F}.
Compact radio
sources are occassionally observed to go through a period of
demagnification at low 
frequencies by roughly a factor of 2;\\
\noindent
(2) pulsar parabolic arcs \citep{2001ApJ...549L..97S}; \\
\noindent
(3) galactic center scattering.

In each of these cases, a simple application of Snell's law with the
assumption of spherical symmetry of the lens requires free electron
densities up to $\sim 10^4$ cm$^{-3}$.  Free electrons are at
temperatures of at least $\sim 10^4$K, and the inferred pressures are
difficult to reconcile with pressure balance in the interstellar
medium.

A solution to case (3) has been proposed by \cite{2006ApJ...640L.159G},
who pointed out that scattering for sheet-like structures is dominated
by the ones most aligned with the line of sight.  The alignment lowers
the required three dimensional electron density.

In this paper, we compute the quantitative consequences of plasma
lenses, and show that triaxial structures are consistent with all
observational data without requiring any unusual physical conditions.
Large axis ratios are generic consequences of reconnection.  In ideal
resistive MHD with ohmic conversion of magnetic fields, current sheets
would be overdense in pressure equilibrium.  Since the resistivity is
almost certainly not ohmic, the actual density is not known.  The
phenomelogy suggests underdense current sheets, which is a probe of
the physics of reconnection.  Geometric factors cause the scattering
to be dominated by rare aligned events.

\section{Plasma Lenses}

\begin{figure}
\centerline{\epsfig{file=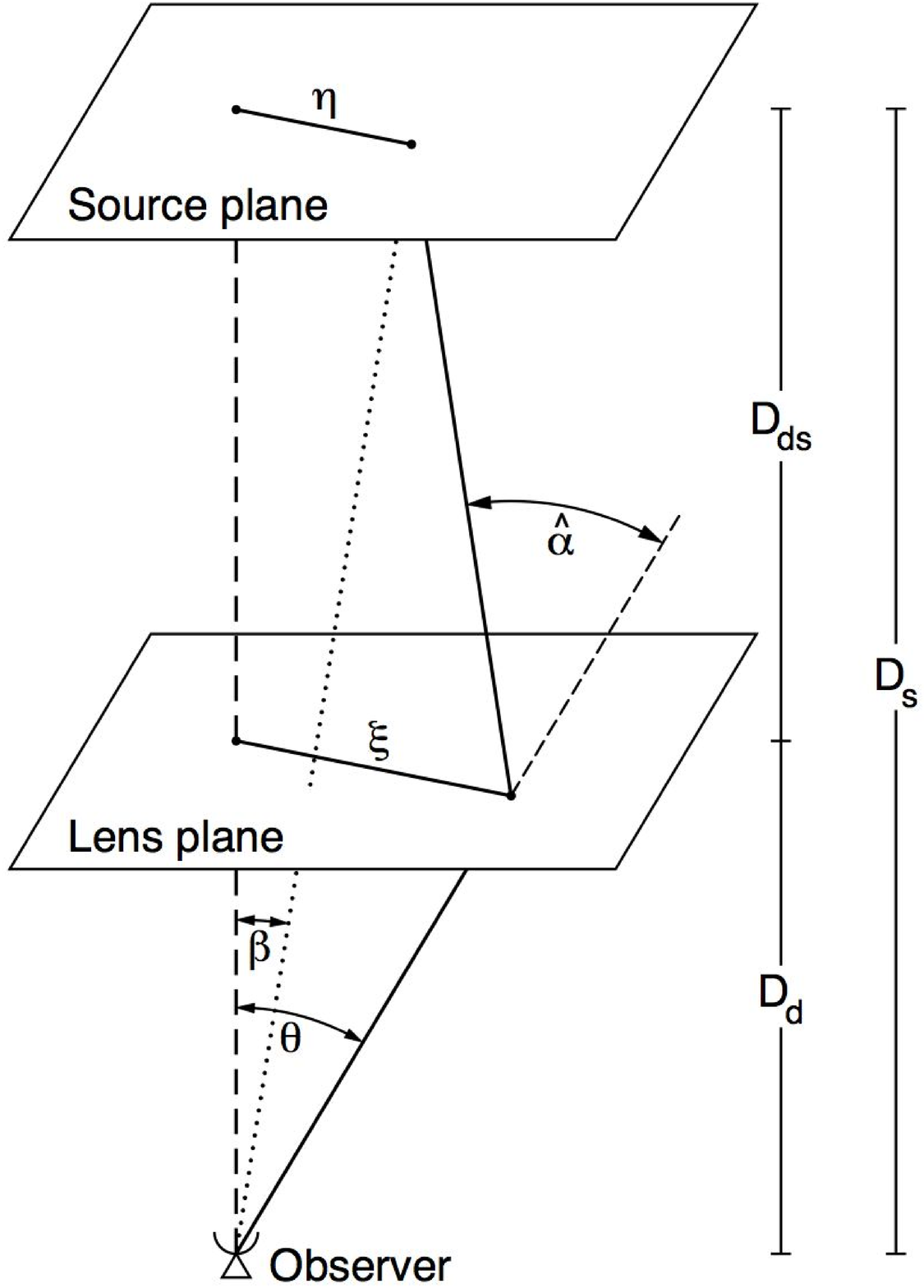, width=3in}}
\caption{Geometry of a lensing event (reproduced from Schneider et al. 1992). 
See text for details.}
\label{fig:lens}
\end{figure}

We follow the notation of \cite{1992grle.book.....S}, reproducing their lensing geometry in 
Fig.\,\ref{fig:lens}. The  diameter distances from the observer to the lens plane and to 
the source plane are ${\rm D}_{\rm d}$ and ${\rm D}_{\rm s}$ respectively, and the 
distance of the source plane from the lens plane is ${\rm D}_{\rm
  ds}$. 
Physical coordinates in the source and lens planes are $\vec\eta$ and $\vec\xi$ respectively, 
defined with respect to the optic axis connecting the observer with the centre 
of the lens. The deflection of a light ray at the lens plane is denoted by $\vec{\hat\alpha}$. 
The angular position of a source at $\vec\eta$ is $\vec\beta$, and $\vec\theta$ is the 
apparent angular position from which the deflected ray travels. They are related through the
lens equation:
\be
\vec\beta  =  \vec\theta-\frac{{\rm D}_{\rm ds}}{{\rm D}_{\rm s}}
\vec{\hat\alpha}({\rm D}_{\rm d}\vec\theta)  =  \vec\theta-\vec\alpha(\vec\theta)
\ee
where $\vec\alpha$ is the deflection angle, simply the difference between the angular position 
of the source and image.

The simplest geometric lens we consider is a triaxial Gaussian
electron density distribution.  This is an extension of 
\citet{1987Natur.328..324R}, where we consider the limit that the axis
ratios are very large, corresponding to thin sheets.  The Romani et al
picture was further quantified by \cite{1998ApJ...496..253C}.  The
main qualitative difference in our new analysis is to consider the
convergent case, which solves several key problems.

In projection, sheets are
highly elongated two dimensional Gaussian surface densities.  The
lensing physics is thus two-dimensional with $\xi$ chosen to be aligned with 
the short axis, with the deflection potential $\psi$ (the projection of the 3-D Newtonian potential) 
given by:
\be
\psi(\theta)=\sigma_\theta^2\kappa_0 \exp(-\theta^2/2\sigma_\theta^2)\,,
\ee
where the gradient of the potential, $\nabla_{\theta}\,\psi$, gives
the deflection angle. 
For a convergent lens, $\kappa_0<0$.

The phase velocity and group velocity of radio waves in a plasma   are 
\be
c_{\rm ph}=c/\sqrt{(1-\omega_{\rm p}^2/\omega^2)}
\ee
and
\be
c_{\rm g}=c\sqrt{(1-\omega_{\rm p}^2/\omega^2)}
\ee
 where
\be
\omega_{\rm p}=\sqrt{n_e e^2/\epsilon_{0}m_e}
\ee
with electron number density $n_{e}$, electron charge and mass $e$ and $m_{e}$ respectively, and $\epsilon_{0}$ being the permittivity of free space. Note that $c_{\rm ph}c_{\rm g}=c^2$.  In the ISM, typical plasma frequencies are kilohertz, so the frequencies of relevance are much larger than the plasma frequency, $\omega\gg\omega_{\rm p}$.
 
The effective refractive index of the plasma and the potential are related through $n =1-2\Phi/c^2 = c/c_{\rm ph}$, 
and thus the intrinsic component of path (or phase) delay, due to
passage through the plasma is frequency-dependent  
$c \,\tau_{\rm grav}=\int \left(n(\omega)-1\right)\,{\rm d}z$.  This
refractive delay is the analogy
to the gravitational Shapiro delay. In the limit $\omega\gg \omega_{\rm p}$,
$\Phi\approx \omega_{\rm p}^2c^2/4\omega^2$.
We use the notation of gravitational lensing, mapping the time delay as the projected potential
\be
\tau = \frac{D_{\rm d}D_{\rm s}}{c\,D_{\rm ds}}\left[\frac{(\vec\theta-\vec\beta)^{2}}{2}-\psi(\vec\theta)\right].
\ee
where the first term accounts for the geometrical delay due to the
offset in source and image positions.

Two possibilities exist: electron overdensities result in a faster
phase velocity, corresponding to a concave (divergent) optical lens.
This case was considered by \cite{1987Natur.328..324R}, which results
in 4 sets of caustics, which are not consistent with the observed
properties of ESEs.  At the center of one of these events, only one image
exists, with flux $\propto 1/\lambda^2$, while
observed events have only order unity flux reduction.  Physically,
electron overdensities can also lead to cooling instabilities, and are
difficult to pressure confine.

Electron underdensities are convergent lenses.  These are generic
consequences of heating processes, for example magnetic reconnection
events.  At the centre of these lenses, as we will demonstrate, 
three images can be seen: a faint central image and two brighter ones on each side of the lens.

The convergence of the lens is related to the laplacian of the
deflection potential, i.e. $\kappa=\psi''/2$ and the components of the
shear for a sheet with the short axis on the plane of the sky are
$\gamma_{1}=\kappa$, $\gamma_{2}=0$.  Thus the magnification
$\mu=1/(1-2\kappa)$.  The convergence is a dimensionless measure of
the distance to the lens in units of its focal lens.  When the
convergence is small, the lensing is weak, and only one image is
formed.

The critical points in the lens plane occur when $1/\mu=0$.  The
previous treatment by \cite{1986ApJ...310..737C} neglected the fact
that the source plane position and the lens plane positions differ,
leading to qualitatively different behaviour in the strong lensing
regime.  In analogy with gravitational lensing, we call the general
lensing geometry ``strong'', to distinguish from the ``weak'' lensing
limit which is easier to compute and has been primarily used in the
scintillation literature.  In weak lensing, the image plane and lens
plane are equivalent, while strong lensing considers the general case.

For the Gaussian lens, we can solve for the critical points and
caustics in the limit of large convergence:
\be
\theta_c=\pm \left(1+\frac{\sqrt{e}}{2\kappa_0}\right), \ \ \ \beta_c = \pm \frac{\kappa_0}{\sqrt{e}}
\ee
Note that here $e\sim 2.718$ refers to the natural log base, not the
electron charge as before.

On axis, the three images for $\beta=0$ appear at $\theta=0, \pm 2\log
(-\kappa_0)$.  Their magnifications are 
\be
\mu=\frac{-1}{1+\kappa_0},\ \frac{1}{2\log(-\kappa_0)-1}.
\label{eqn:mu}
\ee


Note that a convergent lens has three images: two bright and one
faint.  These could be observed with VLBI, or through pulsar
scintillation. 

\begin{figure}
\centerline{\epsfig{file=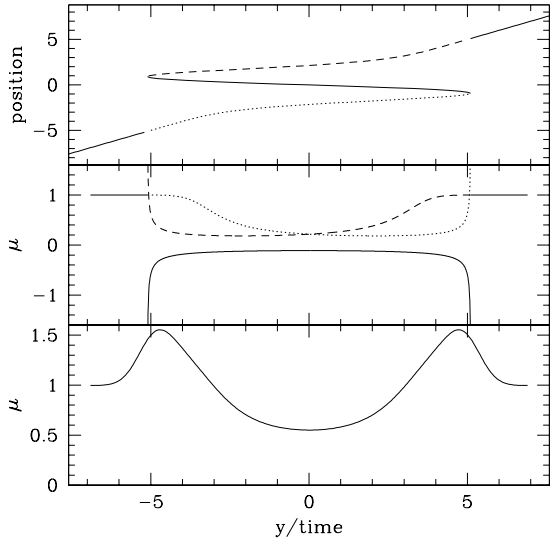, width=3in}}
\caption{Multiple image creation and magnification during an extreme
  scattering event. As a function of source position or time, top
  panel shows positions of image(s), middle and bottom panels show
  their individual and total magnifications respectively.  The bottom
  panel is for a Gaussian source profile of half the angular size of the lens.
Negative
  magnifications correspond to an image parity change, which becomes
  observable in terms of the pulse rotational reflex motion.}

\label{fig:musmooth}
\end{figure}

We show a light curve in Figure~\ref{fig:musmooth}, for a source
moving parallel with respect to the short axis of the sheet. As a
function of source position, or time with respect to direct alignment
between source and lens, the lower panel shows the total image
magnification for a source which is half the size of the lens, the
middle panel shows the magnification of the individual images for a
point source, and the
upper panel shows the corresponding positions of the images (indicated
with the same line styles). Note the production of 3 images and boost
in magnification upon caustic crossing, and the net total
demagnification during a large period of the event.  This is much like
looking at a magnifying glass at distances longer than the focal
length: the central image is inverted, and demagnified.

Along a path with many randomly oriented lenses, the convergence is
dominated by the lenses where the intermediate axis is aligned to the
line of sight, and in projection appear of comparable size to the
short axis.  Since the lenses are rare, a general line of sight may
not contain any such highly aligned lenses.  Nevertheless, the lensing
is dominated by the most aligned scatterer \citep{2006ApJ...640L.159G}.

\section{Pulsars}

The multiple images of the lens can be thought of as an interferometer
on the source. Pulsars generally remain unresolved under such lensing
events, and the images interfere with each other.

The time delay through a lens is $\tau \propto \psi+\alpha^2/2$.  For a
convergent plasma lens, $\psi<0$, resulting in a negative time delay
on axis, i.e. an advanced pulsar arrival.

\begin{figure}
\centerline{\epsfig{file=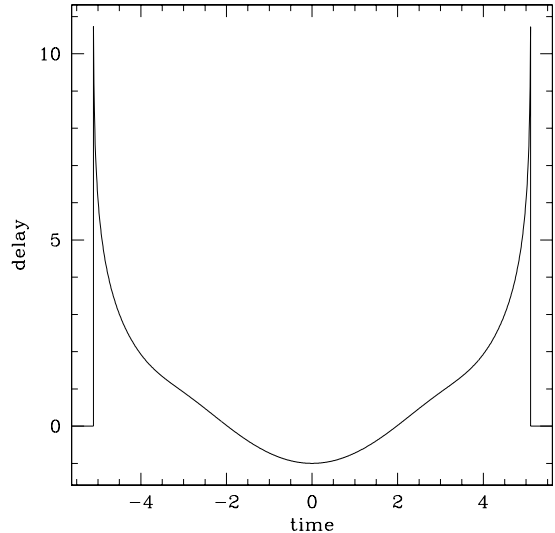, width=3in}}
\caption{Arrival-time delay of pulses from a pulsar averaged over all apparent images.}
\label{fig:pulsedelay}
\end{figure}

Fig.\,\ref{fig:pulsedelay} shows the offset in arrival-time delay of
pulses from a pulsar during a lensing event. Note the sharp rise in
delay near caustic crossing, and the apparent negative delay around
the time when the pulsar and lens are in alignment.  While the phase
velocity in a plasma is larger than the speed of light, the group
velocity is always smaller.  In a convergent plasma lens, the
underdensity results in a lower phase velocity, and thus higher group
velocity, resulting in the apparent advancement in the pulse arrival.

\begin{figure}
\centerline{\epsfig{file=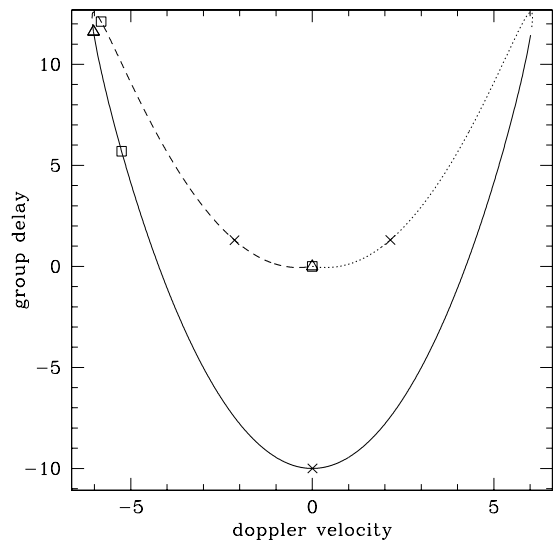, width=3in}}
\caption{Pulsar image trajectory in the secondary spectrum. The
  triangle marks the position at caustic crossing, with one triangle
  superimposed under the central square.  The squares mark the
  positions a short while later, 6\% into the ESE.  The central image
  has barely moved.  The octagons mark
  the image positions at the center of the event.  In a snapshot, only
can only measure relative delays between images.  With pulsar timing,
the actual offset is also observable, leading to a negative group
delay for the central demagnified image.}
\label{fig:delay}
\end{figure}

\subsection{Incidence of lensing}

The incidence of ESE is observed to be roughly 1\% at 2.7 GHz.  At
400 MHz, this gives about a 2\% probability of encountering a lens
with convergence $>$45.  Each lens is visible at 45 radii, implying that
half the time a pulsar dynamic spectrum can show an inverted parabola.

\subsection{Distance measurements}

ESE lensing events in pulsars open up the possibility of precise
geometric distance determinations, which in turn can increase the
sensitivity of gravitational wave detection, and result in high
angular resolution \citep{2010arXiv1010.4337B}.  It requires two
lenses at distinct distances.  During an ESE, two bright images will
have known mutual time delays, readily measurable to a part in 1000 or
better.  There is a reasonable chance of an offset scattering lens at
a different distance: these inverted parabolae are observed in a
substantial fraction of all pulsars.  Using interstellar holography,
the resulting fringe pattern can be decomposed into four distinct
images: two from the ESE lens, and two more from the second scatter.
VLBI allows a measurement of the angular separation of the two ESE
images, and the angular distance to the second scatterer.  These
separations will be comparable to VLBI resolution, yielding fractional
distance errors of $\sim 1/$SN, which can readily exceed $10^3$ for
bright pulsars relevant to gravitational wave timing experiments.  The
effective distance to the ESE is measured. The two images will have
different effective distances to the second scatterer.  The three
distances combined result in a measurement of the three unknowns:
distance to the two lenses, and the distance to the pulsar.

An additional option exists for distance to pulsars in binary systems.
The pulsar performs a periodic orbit which is an ellipse on the sky.
The known orbital parameters, and the time resolved lens measurements,
allow a reconstruction of lens and sources distances.

\section{Extragalactic Radio Sources}

The original ESE were found in extragalactic radio sources.  The
classic example is 0954+658.

At 8.1 GHz, a complex light curve containing multiple peaks was
observed.  At 2.7 GHz, the light curve much resembles the bottom panel
of Figure \ref{fig:musmooth}.  In our picture, we consider that
$\kappa_0 \sim 1$ at 8.1 GHz.  For frequencies close to the critical,
small substructures in the lens or the source can lead to rapid flux
variations, which is observed.  At 2.7 GHz, one then expects $\kappa_0
\sim 10$, which was the value chosen in the figure.  We see that the
model matches the data well.  The factor of two demagnification is a
generic consequence of convergent lenses.  Divergent lenses, on the
other hand, result in a $\sim 1/\kappa_0$ demagnification (see
Eq~\ref{eqn:mu}), which would be an order of magnitude scaling from
the near critical magnfication near 8.1 GHz.

The parallactic alignment of the lens with respect to the source is
not known, making it challenging to directly interpret the existing VLBI
structures of the source.  \citet{2000ApJ...534..706L} noted an
apparent increase in angular size during the ESE, which they
interpreted as inconsistent with refracive lensing, which would
conserve surface brightness and require a decrease in angular size.
In the Gaussian model, the ESE is associated with multiple images with
separations smaller than the VLBI beam.  Thus the apparent size
increase is in fact a generic consequence of this model.

\section{Galactic Center}

The radio source Sgr A* and neighboring masers are observed to be
substantially scatter broadened \citep{1998ApJ...505..715L}, $\theta \sim 1''($GHz/$\nu)^2$.
Models place the screen within $\sim 150$ pc of the galactic center,
implying a scattering angle of order an arcminute.  This has led to
challenges for confining enough plasma to create this scattering.
\cite{2006ApJ...640L.159G} proposed sheets as a solution.
Statistically, the scattering is always dominated by the small
fraction of sheets that are seen face-on.  The convergent lens picture
works in this regime as well.  The recently discovered pulsar
J1746--2850II \citep{2009ApJ...702L.177D} might be behind a lensing
screen, and might exhibit frequency dependent lensing.  Frequent
monitoring would be required.  VLBI measurements of the scattering
disk could also yield insights.  The scattering disk is elongated by
about 2:1 \citep{1993Natur.362...38L}.  This suggests that the number
of scatters is either small, or the magnetic fields have substantial
systematic alignment.  Magnetic field lines must lie in the plane of
the plasma sheets to prevent diffusion and thus lens broadening.
Images are always scattered perpendicular to the field line direction.

\section{Observational Tests}

Two regimes can be searched to confirm this model of these convergent
lenses.  The most direct is the observation of an ESE with VLBI.  It
had been done once by \cite{2000ApJ...534..706L}, but with
insufficient resolution to discern the multiple images.  They note a
slight broadening of the source at the lensing peak, which might be
due to an unresolved mixing of two images.  Lensing conserves surface
brightness, so one expects a larger image to be brighter. The authors
used this argument to rule out a refractive divergent lens. We note,
however, that the expected lens image separation is only modestly, say
a factor of 2 larger than the source.  This results in two images,
each of which is demagnified and smaller, but next to each other to
make them appear more extended.  Their synthesized beam is much larger
than the source size or the separation, so these two effects are not
separable.

Unfortunately, the observational monitoring programme of bright
compact sources has been discontinued.

Pulsars also undergo extreme scattering events
\citep{1993Natur.366..320C}.  In such cases, the three images will
coherently interfere with each other, resulting in a distinctive
signature in the secondary spectrum, shown in Figure \ref{fig:delay}.
This figure plots the raw delays which can be obtained from
interstellar holography\citep{2008MNRAS.388.1214W}.

\section{Physical Discussion}

So far, the discussion has focussed on the quantitative properties of
the convergent lens.  Some comments on its relation to the physical
properties of the lenses are in order.  

Typical electron densities as measured by pulsar dispersion measure
are $n_e \sim 0.03$.  This gives a phase velocity about $10^{-12}$
larger than $c$ at frequencies of $\sim$ GHz. For an underdense lens,
the maximal deflection angle from Snell's law is therefore
$10^{-4}/\alpha$ mas, where $\alpha$ is the angle between the lens
surface and the line of sight.  Grazing incidence with $\alpha \sim
10^{-4}$ is needed to explain the observed refractive angles of $\sim$
mas. 

We speculate that reconnection current sheets could satisfy the
properties required by the phenomenology.  Unfortunately, the
quantitative process of reconnection is very poorly understood.  The
thickness of a current sheet is probably related to resistivity.  The
primary quantitative calculable ohmic restivity leads to very long
reconnection time scales, leading to problems with the generation of
magnetic fields.  Alternative reconnection processes, or fast
reconnection, has been proposed.  The reconnection rate and sheet
geometry differs by many orders of magnitude between proposed
scenarios \citep{2010PhPl...17j2302P}.  Without a known reconnection
rate, the incidence rate of current sheets is not predictable, nor is
the lens thickness. Similarly, the aspect ratio of the sheets is not
known.  One qualitatively expects the longest axis to be related to
the curvature of field lines, which determines the amount of area that
approaches sufficiently closely to undergo reconnection.  The shortest
axis is given by the unknown resistive scale, which could be as short
as the electron gyromagnetic radius.  To explain the phenomenology of
ESE, we require the ratio of the intermediate axis to the short axis
to be $\gtrsim 10^4$. Unfortunately, the statistical properties of
ESEs is not predictable within our understanding of plasma physics.
Instead, we can use the properties of the ESE to learn about
reconnection.

\section{Conclusions}

We have introduced triaxial convergent plasma lenses to explain three
separate, but related, refractive plasma lensing effects. Reconnection
events generically lead to highly flattened current sheets.  We have
found that generic physical conditions can account for all the known
phenomena, which are otherwise challenging to explain.  These include
the apparent distance localization of scatters, the modest frequency
dependence of ESE, overpressure problem.  It also suggests that the
apparent diffractive phenomena in pulsar may be due to interference of
refractive images, rather than stochastic fresnel scale lensing.

\section{Acknowledgements}

U-LP thanks NSERC and the KICC for support and hosting the visit which
lead to this paper, and LJK thanks the Royal Society.  We would like
to thank Peter Goldreich, Barney Rickett, Ethan Vishniac and Chris
Thompson for helpful discussions.

\newcommand{\araa}{ARA\&A}   
\newcommand{\afz}{Afz}       
\newcommand{\aj}{AJ}         
\newcommand{\azh}{AZh}       
\newcommand{\aaa}{A\&A}      
\newcommand{\aas}{A\&AS}     
\newcommand{\aar}{A\&AR}     
\newcommand{\apj}{ApJ}       
\newcommand{\apjs}{ApJS}     
\newcommand{\apjl}{ApJ}      
\newcommand{\apss}{Ap\&SS}   
\newcommand{\baas}{BAAS}     
\newcommand{\jaa}{JA\&A}     
\newcommand{\mnras}{MNRAS}   
\newcommand{\nat}{Nat}       
\newcommand{\pasj}{PASJ}     
\newcommand{\pasp}{PASP}     
\newcommand{\paspc}{PASPC}   
\newcommand{\qjras}{QJRAS}   
\newcommand{\sci}{Sci}       
\newcommand{\sova}{SvA}      
\newcommand{\aap}{A\&A}

\bibliography{ese}
\bibliographystyle{mn2e}

\label{lastpage}

\end{document}